\documentstyle[12pt]{article}
\newcommand{\beq}{\begin{equation}}
\newcommand{\eeq}{\end{equation}}

\begin{document}

\begin{center}
{\bf  Magnetic field-induced Landau Fermi Liquid in high-$T_c$
metals }\bigskip

{M.Ya. Amusia$^{a,b}$, V.R.
Shaginyan$^{a,c}$ \footnote{E--mail: vrshag@thd.pnpi.spb.ru}}\\
\bigskip
{\it $^{a\,}$The Racah Institute of Physics,
the Hebrew University, Jerusalem 91904, Israel;\\
$^{b\,}$A.F. Ioffe Physical-Technical Institute, 194021 St.
Petersburg, Russia;\\
$^{c\,}$Petersburg Nuclear Physics Institute, Gatchina, 188300,
Russia}

\end{center}

\begin{abstract}

We consider the behavior of strongly correlated electron liquid in
high-temperature superconductors within the framework of the
fermion condensation model. We show that at low temperatures the
normal state recovered by the application of a magnetic field
larger than the critical field  can be viewed as the Landau Fermi
liquid induced by the magnetic field. In this state, the
Wiedemann-Franz law and the Korringa law are held and the elementary
excitations are the Landau Fermi Liquid quasiparticles.
Contrary to what might be expected from the Landau theory,
the effective mass of quasiparticles depends on the magnetic field.
The recent experimental verifications of the Wiedemann-Franz law in
heavily hole-overdoped, overdoped and optimally doped cuprates  and
the verification of the Korringa law in the electron-doped
copper-oxide superconductor strongly support the existence of fermion
condensate in high-$T_c$ metals.

\end{abstract}\bigskip



\newpage

The Landau Fermi Liquid (LFL) theory belongs to the most important
achievements in physics \cite{lan}. The LFL theory has revealed
that the low-energy elementary excitations of a Fermi liquid look
like the spectrum of an ideal Fermi gas. These excitations are
described in terms of quasiparticles with an effective mass $M^*$,
charge $e$ and spin $1/2$. The quasiparticles define the major
part of the low-temperature properties of Fermi liquids. Note that
this powerful theory gives theoretical grounds for the theory of
conventional superconductivity. On the other hand, despite
outstanding results obtained in the condensed matter physics, the
understanding  of the rich and striking behavior of the
high-temperature superconductors remains, as years before, among
the main problems of the condensed matter physics. There is also a
fundamental question about whether or not the properties of this
electron liquid in high-$T_c$ metals can be understood within the
framework of the LFL theory.

It was reported recently that in the normal state obtained by
applying a magnetic field greater than the upper critical filed
$B_c$, in a hole-doped cuprates at overdoped concentration
(Tl$_2$Ba$_2$CuO$_{6+\delta}$) \cite{cyr} and at optimal doping
concentration (Bi$_2$Sr$_2$CuO$_{6+\delta}$) \cite{cyr1}, there
are no any sizable violations of the Wiedemann-Franz (WF) law.
In the electron-doped copper oxide superconductor
Pr$_{0.91}$LaCe$_{0.09}$Cu0$_{4-y}$ ($T_c$=24 K) when
superconductivity is removed by a strong magnetic field,
it was found that the spin-lattice relaxation rate $1/T_1$
follows the $T_1 T=constant$ relation, known as Korringa law
\cite{korr}, down to temperature of $T=0.2$ K \cite{zheng}. At
elevated temperatures and applied magnetic fields of 15.3 T
perpendicular to the CuO$_2$ plane, $1/T_1T$ as a function of $T$ is
a constant below $T=55$ K.
At $300$ K $>T>50$ K, $1/T_1T$  decreases with
increasing $T$ \cite{zheng}. Recent measurements for strongly
overdoped non-superconducting La$_{1.7}$Sr$_{0.3}$CuO$_4$ have shown
that the resistivity $\rho$ exhibits $T^2$ behavior,
$\rho=\rho_0+\Delta\rho$ with $\Delta\rho=AT^2$, and the WF law is
verified to hold perfectly \cite{nakam}. Since the validity of the WF
law and of the Korringa law are a robust signature of LFL, these
experimental facts demonstrate that the observed elementary
excitations cannot be distinguished from the Landau quasiparticles.
This imposes strong constraints for models describing the hole-doped
and electron-doped high-temperature superconductors. For example, in
the cases of a Luttinger liquid \cite{kane}, spin-charge separation
(see e.g.  \cite{sen}), and in some solutions of $t-J$ model
\cite{hough} a violation of the WF law was predicted.

In this Letter, we consider the behavior of strongly correlated
electron liquid in high-temperature metals within the framework of
the fermion condensation model \cite{ks,ksk,ms}. We show that at
temperatures $T\to 0$ the normal state recovered by the
application of a magnetic field larger than the critical field
$B_c$ can be viewed as  LFL induced by the magnetic field. In this
state, the WF law and the Korringa law are held and the elementary
excitations are LFL quasiparticles.
We show that contrary to what might be expected from the Landau
theory, the effective mass of quasiparticles depends on the magnetic
field.

At $T\to 0$, the ground state energy $E_{gs}[\kappa({\bf p}),n({\bf
p})]$ of an electron liquid is a functional of the order parameter
of the superconducting state $\kappa({\bf p})$ and of the
quasiparticle occupation numbers $n({\bf p})$. Here we assume that
the electron system is two-dimensional, while all results can be
transported to the case of three-dimensional system. This energy
is determined by the known equation of the weak-coupling theory of
superconductivity, see e.g. \cite{til} \beq E_{gs}\ =\ E[n({\bf
p})]+\int \lambda_0V({\bf p}_1,{\bf p}_2) \kappa({\bf p}_1)
\kappa^*({\bf p}_2) \frac{d{\bf p}_1d{\bf p}_2}{(2\pi)^4}\ . \eeq
Here $E[n({\bf p})]$ is the ground-state energy of a normal Fermi
liquid, $n({\bf p})=v^2({\bf p})$ and $\kappa({\bf p})=v({\bf
p})\sqrt{1-v^2({\bf p})}$. It is assumed that the pairing
interaction $\lambda_0V({\bf p}_1,{\bf p}_2)$ produced, for
instance, by electron-phonon interaction in high-$T_c$ metals is
weak. Minimizing $E_{gs}$ with respect to $\kappa({\bf p})$ we
obtain the equation connecting the single-particle energy
$\varepsilon({\bf p})$ to the superconducting gap $\Delta({\bf
p})$, \beq \varepsilon({\bf p})-\mu\ =\ \Delta({\bf p})
\frac{1-2v^2({\bf p})} {2\kappa({\bf p})}, \eeq where the
$\mu$ is the chemical potential, and
single-particle energy $\varepsilon({\bf p})$ is determined by the
Landau equation \beq \varepsilon({\bf p})= \frac{\delta E[n({\bf
p})]}{\delta n({\bf p})}. \eeq Note that $E[n({\bf p})]$,
$\varepsilon[n({\bf p})]$, and the Landau amplitude $F_L({\bf
p},{\bf p}_1)=\delta E^2[n({\bf p})]/\delta n({\bf p})\delta({\bf
p}_1)$ implicitly depend on the density $x$ which defines the
strength of $F_L$. The equation for the superconducting gap
$\Delta({\bf p})$ is of the form
\begin{eqnarray} \Delta({\bf p}) &=& -\int\lambda_0V({\bf p},{\bf
p}_1)\kappa({\bf p}_1)
\frac{d{\bf p}_1}{4\pi^2} \nonumber\\
&=& -\frac{1}{2}\int\lambda_0 V({\bf p},{\bf p}_1)
\frac{\Delta({\bf p}_1)}{\sqrt{(\varepsilon({\bf p}_1)-\mu)^2
+\Delta^2({\bf p}_1)}} \frac{d{\bf p}_1}{4\pi^2}\ .
\end{eqnarray}
If $\lambda_0\to 0$, than the maximum value of the superconducting
gap $\Delta_1\to 0$, and Eq. (2) reduces to the equation
\cite{ks,ms,ams} \beq \varepsilon({\bf p})-\mu\ =\ 0,\quad \mbox{
if}\quad 0<n({\bf p})<1;\: p_i\leq p\leq p_f\ . \eeq A new state
of electron liquid with the fermion condensate (FC) \cite{ks,vol}
is defined by Eq. (5) and characterized by a flat part of the
spectrum in the $(p_i-p_f)$ region. Apparently, the momenta $p_i$
and $p_f$ have to satisfy $p_i<p_F<p_f$, where $p_F$ is the Fermi
momentum. When the amplitude $F_L(p=p_F,p_1=p_F)$ as a function of
the density $x$ becomes sufficiently small, the flat part
vanishes, Eq. (5) has the only trivial solution
$\varepsilon(p=p_F)=\mu$, and the quasiparticle occupation numbers
are given by the step function, $n({\bf p})= \theta(p_F-p)$
\cite{ks,ksk}. Note, that a formation of the flat part of the
spectrum has been recently confirmed in Ref. \cite{dzyal,irkh}.

Now we can study the relationships between the state defined by
Eq. (5) and the superconductivity. At $T\to 0$, Eq. (5) defines a
particular state of a Fermi liquid with FC, for which the modulus
of the order parameter $|\kappa({\bf p})|$ has finite values in
the $(p_i-p_f)$ region, whereas  $\Delta_1\to 0$ unlike
the conventional theory of superconductivity, which demands that
$\Delta_1$ vanishes simultaneously with the order parameter
$\kappa({\bf p})$. Such a situation can take place because the
order parameter $\kappa({\bf p})$, as it follows from Eqs. (3) and
(5), is determined by the Landau amplitude $F_L$, while $\Delta_1$ is
given by Eq. (4) and tends to zero at $\lambda_0\to 0$.  This state
can be considered as superconducting, with an infinitely small value
of $\Delta_1$, so that the entropy of this state is equal to zero. It
is obvious that this state being driven by the quantum phase
transition disappears at $T>0$ \cite{ms}.  Any quantum phase
transition, which takes place at temperature $T=0$, is determined by
a control parameter other than temperature, for example, by pressure,
by magnetic field, or by the density of mobile charge carriers $x$.
The quantum phase transition occurs at a quantum critical point. At
some density $x\to x_{FC}$, when the Landau amplitude $F_L$ becomes
sufficiently weak, and $p_i\to p_F\to p_f$, Eq. (5) determines the
critical density $x_{FC}$ at which the fermion condensation quantum
phase transition (FCQPT) takes place leading to the formation of FC
\cite{ks,ms}. It follows from Eq. (5) that the system becomes divided
into two quasiparticle subsystems:  the first subsystem in the
$(p_i-p_f)$ range is characterized  by the quasiparticles with the
effective mass $M^*_{FC}\propto 1/\Delta_1$, while the second one is
occupied by quasiparticles with finite mass $M^*_L$ and momenta
$p<p_i$. The density of states near the Fermi level tends to
infinity, $N(0)\propto M^*_{FC}\propto 1/\Delta_1$ \cite{ms}.

If $\lambda_0\neq 0$, then $\Delta_1$ becomes finite. It is seen
from Eq. (4) that the superconducting gap depends on the
single-particle spectrum $\varepsilon({\bf p})$. On the other
hand, it follows from Eq. (2) that $\varepsilon({\bf p})$ depends
on $\Delta({\bf p})$ provided that at $\Delta_1\to 0$ the spectrum
is given by Eq. (5). Let us assume that $\lambda_0$ is small so
that the particle-particle interaction $\lambda_0 V({\bf p},{\bf
p}_1)$ can only lead to a small perturbation of the order
parameter $\kappa({\bf p})$ determined by Eq. (5). Upon
differentiation both parts of Eq. (2) with respect to the momentum
$p$, we obtain that the effective mass
$M_{FC}=d\varepsilon(p)/dp_{\,|p=p_F}$ becomes finite \cite{ms}
\beq M^*_{FC}\sim p_F\frac{p_f-p_i}{2\Delta_1}. \eeq It is clear
from Eq. (6) that the effective mass and the density of states
$N(0)\propto M^*_{FC}\propto 1/\Delta_1$ are finite. As a result,
we are led to the conclusion that in contrast to the conventional
theory of superconductivity the single-particle spectrum
$\varepsilon({\bf p})$ strongly depends on the superconducting gap
and we have to solve Eqs. (3) and (4) in a self-consistent way.

Since the particle-particle interaction is small, the order
parameter $\kappa({\bf p})$ is governed mainly by FC and we can
solve Eq. (4) analytically  taking the Bardeen-Cooper-Schrieffer
approximation for the particle-particle interaction:
$\lambda_0V({\bf p},{\bf p}_1)=-\lambda_0$ if $|\varepsilon({\bf
p})-\mu|\leq \omega_D$, i.e. the interaction is zero outside this
region, with $\omega_D$ being the characteristic phonon energy. As
a result, the maximum value of the superconducting gap is given by
\cite{ams} \beq \Delta_1(0)\ \simeq \frac{\lambda_0
p_F(p_f-p_F)}{2\pi}\ln\left(1+\sqrt2\right) \simeq
2\beta\varepsilon_F \frac{p_f-p_F}{p_F}\ln\left(1+\sqrt2\right).
\eeq Here, the Fermi energy $\varepsilon_F=p_F^2/2M^*_L$, and the
dimensionless coupling constant $\beta$ is given by the relation
$\beta=\lambda_0 M^*_L/2\pi$. Taking the usual values of $\beta$
as $\beta\simeq 0.3$, and assuming $(p_f-p_F)/p_F\simeq 0.2$, we
get from Eq. (7) a large value of $\Delta_1(0)\sim
0.1\varepsilon_F$, while for normal metals one has
$\Delta_1(0)\sim 10^{-3}\varepsilon_F$. Now we determine the
energy scale $E_0$ which defines the region occupied by
quasiparticles with the effective mass $M^*_{FC}$ \beq E_0=
\varepsilon({\bf p}_f)-\varepsilon({\bf p}_i) \simeq 2
\frac{(p_f-p_F)p_F}{M^*_{FC}}\ \simeq\ 2\Delta_1. \eeq

We have returned back to the Landau Fermi liquid theory since high
energy degrees of freedom are eliminated and the quasiparticles
are introduced. The only difference between LFL, which serves as a
basis when constructing the superconducting state, and Fermi
liquid after FCQPT is that we have to expand the number of
relevant low energy degrees of freedom by introducing new type of
quasiparticles with the effective mass $M^*_{FC}$ given by Eq. (6)
and the energy scale $E_0$ given by Eq. (8). Therefore, the
single-particle spectrum $\varepsilon({\bf p})$ of system in
question is characterized by two effective masses $M^*_L$ and
$M^*_{FC}$ and by the scale $E_0$, which define the low
temperature properties including the line shape of quasiparticle
excitations \cite{ms,ams}. We note that both the effective mass
$M^*_{FC}$ and the scale $E_0$ are temperature independent at
$T<T_c$, where $T_c$ is the critical temperature of the
superconducting phase transition \cite{ams}. Obviously, we cannot
directly relate these new quasiparticle excitations with the
quasiparticle excitations of an ideal Fermi gas because the system
in question has undergone FCQPT. Nonetheless, the main basis of
the Landau Fermi liquid theory survives FCQPT: the low energy
excitations of a strongly correlated liquid with FC are
quasiparticles.

As it was shown above, properties of these new quasiparticles are
closely related to the properties of the superconducting state. We
may say that the quasiparticle system in the range $(p_i-p_f)$
becomes very ``soft'' and is to be considered as a strongly
correlated liquid. On the other hand, the system's properties and
dynamics are dominated by a strong collective effect having its
origin in FCQPT and determined by the macroscopic number of
quasiparticles in the range $(p_i-p_f)$. Such a system cannot be
perturbed by the scattering of individual quasiparticles and has
features of a ``quantum protectorate" \cite{ms,rlp,pa}.

As any phase transition, FCQPT is related to the order parameter,
which induces a broken symmetry. As we have seen above, the order
parameter is the superconducting order parameter $\kappa({\bf
p})$, while $\Delta_1$ being proportional to the coupling constant
(see Eqs. (5) and (7)) can be small. Therefore, the existence of
such a state, that is electron liquid with FC, can be revealed
experimentally. Since the order parameter $\kappa({\bf p})$ is
suppressed by the critical magnetic field $B_c$, when $B_c^2\sim
\Delta_1^2$. If the coupling constant $\lambda_0\to 0$, the weak
critical magnetic field $B_c\to 0$ will destroy the state with FC
converting the strongly correlated Fermi liquid into  LFL. In this
case the magnetic field play a role of the control parameter
determining the effective mass \cite{pogsh}
\begin{equation}
M^*_{FC}(B)\propto \frac{1}{\sqrt{B}}.
\end{equation}
Note, that the outlined
above behavior was observed experimentally in the heavy-electron
metal YbRh$_2$Si$_2$ \cite{gen}.

If $\lambda_0$ is finite, the critical
field is also finite, and Eq. (9) is valid at $B>B_c$. In that
case, the effective mass $M^*_{FC}(B)$ is finite,
and the system is driven back to LFL and
has the LFL behavior induced by the magnetic field.
At a constant magnetic field, the low energy elementary excitations
are characterized by $M^*_{FC}(B)$ and cannot be distinguished from
Landau quasiparticles. As a result, at $T\to 0$, the WF law is held
in accordance with experimental facts \cite{cyr,cyr1}.  On the hand,
in contrast to the LFL theory, the effective mass  $M^*_{FC}(B)$
depends on the magnetic field.

Equation (9) shows that by applying a magnetic field $B>B_c$ the
system can be driven back into LFL with the effective mass
$M^*_{FC}(B)$ which is finite and independent of the temperature.
This means that the low temperature properties depend on the
effective mass in accordance with the LFL theory. In particular,
the resistivity $\rho(T)$ as a function of the temperature behaves
as $\rho(T)=\rho_0+\Delta\rho(T)$ with $\Delta\rho(T)=AT^2$, and the
factor $A\propto (M^*_{FC}(B))^2$.  At finite temperatures, the system
persists to be LFL, but there is the crossover from the LFL
behavior to the non-Fermi liquid behavior at temperature
$T^*(B)\propto \sqrt{B}$. At $T>T^*(B)$, the effective mass
starts to depend on the temperature $M_{FC}\propto 1/T$, and the
resistivity possesses the non-Fermi liquid behavior with a
substantial linear term, $\Delta\rho(T)=aT+bT^2$ \cite{ms,pogsh}.
Such a behavior of the resistivity was observed in the
cuprate superconductor Tl$_2$Ba$_2$CuO$_{6+\delta}$
($T_c<$ 15 K) \cite{mack}. At B=10 T, $\Delta\rho(T)$ is a
linear function of the temperature
between 120 mK and 1.2 K, whereas at B=18 T, the temperature
dependence of the resistivity is consistent with
$\rho(T)=\rho_0+AT^2$ over the same temperature range \cite{mack}.

In LFL, the nuclear spin-lattice relaxation rate $1/T_1$
is determined by the quasiparticles near the Fermi level whose
population is proportional to $M^* T$, so that $1/T_1T\propto M^*$
is a constant \cite{zheng,korr}. As it was shown, when the
superconducting state is removed by the application of a magnetic
field, the underlying ground state can be seen as the field induced
LFL with effective mass depending on the magnetic field.  As a
result, the rate $1/T_1$ follows the $T_1T=constant$ relation, that
is the Korringa law is held.  Unlike the behavior of LFL, as it
follows from Eq. (9), $1/T_1T\propto M^*_{FC}(B)$ decreases with
increasing the magnetic field at $T<T^*(B)$.  While, at $T>T^*(B)$,
we observe that $1/T_1T$ is a decreasing function of the temperature,
$1/T_1T\propto M^*_{FC}\propto 1/T$.  These observations are in a
good agreement with the experimental facts \cite{zheng}.  Since
$T^*(B)$ is an increasing function of the magnetic field, the
Korringa law retains its validity to higher temperatures at elevated
magnetic fields. We can also conclude, that at temperature $T_0\leq
T^*(B_0)$ and elevated magnetic fields $B>B_0$, the system shows a
more pronounced metallic behavior since the effective mass decreases
with increasing $B$, see Eq. (9). Such a behavior of the effective
mass can be observed in the de Haas van Alphen-Shubnikov studies,
$1/T_1T$ measurements, and the resistivity measurements. These
experiments can shed light on the physics of high-$T_c$ metals and
reveal relationships between high-$T_c$ metals and heavy-electron
metals.

The existence of FCQPT can also be revealed experimentally because
at densities $x>x_{FC}$, or beyond the FCQPT point, the system
should be LFL at sufficiently low temperatures \cite{shag}. Recent
experimental data have shown that this liquid exists in heavily
overdoped non-superconducting La$_{1.7}$Sr$_{0.3}$CuO$_4$
\cite{nakam}. It is remarkable that up to $T=55$ K the resistivity
exhibits the $T^2$ behavior with no additional linear term, and the
WF law is verified to within the experimental resolution
\cite{nakam}. While at elevated temperatures, a strong deviations
from the LFL behavior are observed. We anticipate that the system can
be again driven back to the LFL behavior by the application of
sufficiently strong magnetic fields \cite{shag}.

In summary, we have shown that the Landau Fermi liquid properties
of high-$T_c$ metals observed at low temperatures in
optimally hole-doped, overdoped and electron-doped cuprates by the
application of a magnetic field higher than the critical field can be
explained within the frameworks of the fermion condensation theory of
high-$T_c$ superconductivity.  In that case of LFL obtained by
applying magnetic field, the effective mass depends on the
magnetic field, the WF law and the Korringa law are held. These
observations are in good agreement with experimental facts. The
recent experimental verifications of the WF law in heavily overdoped,
overdoped and optimally doped cuprates and the verification of the
Korringa law in the electron-doped copper-oxide superconductor give
strong support in favor of the existence of FC in high-$T_c$ metals.

MYA is grateful to the Hebrew University Intramural fund of the
Hebrew University for financial support. VRS is grateful to Racah
Institute of Physics for the hospitality during his stay in
Jerusalem. This work was supported in part by the Russian
Foundation for Basic Research, project No 01-02-17189.

\end{document}